\documentclass[12pt]{article}
\usepackage{amsmath,amsfonts, amssymb, braket, bbold}
\usepackage{textpos,enumitem,marginnote}
\usepackage{tikz}
\usetikzlibrary{trees,er,snakes,shapes,mindmap}
\usepackage{titlesec}
\titleformat{\section}
 {\normalfont\fontfamily{put}\fontsize{13pt}{16pt}\bfseries\color{black}}
{\thesection}{1em}{}
\titleformat{\subsection}
 {\normalfont\fontfamily{put}\fontsize{12pt}{16pt}\bfseries\color{black}}
{\thesubsection}{1em}{}

\def \beq  {\begin{equation}}
\def \eeq  {\end{equation}}
\def \beqar {\begin{eqnarray}}
\def \eeqar {\end{eqnarray}}
\allowdisplaybreaks
\def\sqr#1#2{{\vcenter{\vbox{\hrule height.#2pt
\hbox{\vrule width.#2pt height#1pt \kern#1pt
\vrule width.#2pt}\hrule height.#2pt}}}}

\def\Tr {{\rm Tr}}

\def\del {\partial}

\def\A {{\cal A}}
\def\B {{\cal B}}

\def\E {{\cal E}}

\def\G {{\cal G}}

\def\M{{\cal M}}
\def\O {{\cal O}}

\def\half{\textstyle{1\over 2}}

\mathchardef\mhyphen="2D
\begin{document}
%%%%%%%%%%%%%%%%%%%%%%%%%%%%%%%%%%%%%%%%%%%%%%%
%%%%%%%%%%%%%%%%%%%%%%%%%%%%%%%%%%%%%%%%%%%%%%%
\fontfamily{bch}\fontsize{12pt}{16pt}\selectfont
%\fontfamily{pnb}\fontsize{12pt}{16pt}\selectfont
%\fontfamily{pzc}\fontsize{14pt}{16pt}\selectfont
%\fontfamily{pbk}\fontsize{12pt}{16pt}\selectfont
%\fontfamily{cmr}\fontsize{11pt}{15pt}\selectfont
%\fontfamily{put}\fontsize{12pt}{17pt}\selectfont
%\fontfamily{lmss}\fontsize{11pt}{16pt}\selectfont
%\fontfamily{phv}\fontshape{ro}\fontsize{11pt}{14pt}\selectfont
%\fontfamily{ptm}\fontseries{m}\fontshape{r}\fontsize{12pt}{16pt}\selectfont
%\fontfamily{pnc}\fontseries{m}\fontshape{r}\fontsize{11pt}{15pt}\selectfont
%\fontfamily{ppl}\fontseries{m}\fontshape{r}\fontsize{11pt}{15pt}\selectfont
%\usefont{T1}{phv}{m}{it}
%%%%%%%%%%%%%%%%%%%%%%%%%%%%%%%%%%%%%%%%%%%%%%%
%%%%%%%%%%%%%%%%%%%%%%%%%%%%%%%%%%%%%%%%%%%%%%%
%%%%%%%%%%%%%%%%%%%%%%%%%%%%%%%%%%%%%%%%%%%%%%%
%%%%%%%%%%%%%%%%%%%%%%%%%%%%%%%%%%%%%%%%%%%%%%%
\def \CMP {{Commun. Math. Phys.}}
\def \PRL {{Phys. Rev. Lett.}}
\def \PL {{Phys. Lett.}}
\def \NPBProc {{Nucl. Phys. B (Proc. Suppl.)}}
\def \NP {{Nucl. Phys.}}
\def \RMP {{Rev. Mod. Phys.}}
\def \JGP {{J. Geom. Phys.}}
\def \CQG {{Class. Quant. Grav.}}
\def \MPL {{Mod. Phys. Lett.}}
\def \IJMP {{ Int. J. Mod. Phys.}}
\def \JHEP {{JHEP}}
\def \PR {{Phys. Rev.}}
\def \JMP {{J. Math. Phys.}}
\def \GRG{{Gen. Rel. Grav.}}
%%%%%%%%%%%%%%%%%%%%%%%%%%%%%%%%%%%%%%%%%%%%%%%
%%%%%%%%%%%%%%%%%%%%%%%%%%%%%%%%%%%%%%%%%%%%%%%
%%%%%%%%%%%%%%%%%%%%%%%%%%%%%%%%%%%%%%%%%%%%%%%
%%%%%%%%%%%%%%%%%%%%%%%%%%%%%%%%%%%%%%%%%%%%%%%
%%%%%%%%%%%%%%%%%%%%%%%%%%%%%%%%%%%%%%%%%%%%%%%
\begin{titlepage}
\title{Superselection, Boundary Algebras and Duality in Gauge Theories}

\author{A.~P. Balachandran$^1$\footnote{balachandran38@gmail.com}, V.~P. Nair$^2$\footnote{vpnair@ccny.cuny.edu}, A. Pinzul$^{3,4}$\footnote{aleksandr.pinzul@gmail.com}, A.~F. Reyes-Lega$^5$\footnote{anreyes@uniandes.edu.co}\, \\ and S. Vaidya$^6$\footnote{vaidya@iisc.ac.in}\\
${}^1${\small Department of Physics, Syracuse University, Syracuse, New York 13244-1130, USA}\\
$^2${\small Physics Department, City College of the CUNY, New York, New York 10031, USA}\\
$^3${\small Universidade de Brasília, Instituto de Física 70910-900, Brasília, DF, Brazil}\\
$^4${\small International Center of Physics C.P. 04667, Brasília, DF, Brazil}\\
${}^5${\small Departamento de F\'{i}sica, Universidad de los Andes,  A.A. 4976-12340, Bogot\'a, Colombia}\\
${}^6${\small Centre for High Energy Physics,  Indian Institute of Science, Bengaluru, 560012, India}\\
}

\date{}

\maketitle
\begin{abstract}
We consider the generators of gauge transformations with test functions which do not vanish on the boundary of a spacelike region of interest.
These are known to generate the edge degrees of freedom in a gauge theory.
In this paper, we augment these by introducing
the dual or magnetic analogue of such operators.
We then study the algebra
of these operators, focusing on implications for the
superselection sectors of the gauge theory. 
 A manifestly duality-invariant action is also
considered, from which alternate descriptions which are $SL(2, \mathbb{Z})$ transforms of each other can be obtained.
We also comment on a number of issues related to local charges, definition of confinement and the appearance of interesting mathematical structures such as the Drinfel'd double and the Manin triple.

\end{abstract}

\end{titlepage}
%%%%%%%%%%%%%%%%%%%%%%%%%%%%%%%%%%%%%%%%%%%%%%%
%%%%%%%%%%%%%%%%%%%%%%%%%%%%%%%%%%%%%%%%%%%%%%%
\pagestyle{plain} \setcounter{page}{2}
\section{Introduction}

Superselection sectors are a characteristic feature of quantum field theories  
owing to the presence of an infinite number of degrees of freedom. Their effects are particularly significant  and manifest in gauge theories such as QED or QCD, where they can strongly constrain the nature of observables.
Thus in QED one has the constraint of the Gauss law in the sense that all observables are required to commute with it. The charge operator, which is closely related to the Gauss law, commutes with it, and with all local observables as well. In other words, the charge algebra  is a nontrivial commutant of the local algebra of observables. This is a fundamental difference between gauge theories and field theories without gauge symmetry.

But there are further implications of this framework,
even for the abelian gauge group of QED. Let us recall the meaning of the gauge group ${\cal G}(G)$ based on a compact Lie group $G$;
the latter may be referred to as the `global' group. Consider fields defined on an $N$-dimensional Minkowski space $M^N$.
In this case, ${\cal G}(G)$ is generally isomorphic to the group of maps from the spatial `boundary' of the spatial slice $M^{N-1}$, which is taken to be $S^{N-2}$, to $G$. They are generated by the Gauss law operator appropriately smeared with test functions. A moderately careful treatment of these test functions smearing the Gauss law shows that there is an infinity of
 superselected operators \cite{Balachandran:2013wsa,Balachandran:2018cgw}. These generate what we have previously referred to as the sky group; effectively, they measure the moments of the electric fields at infinity. In section 2, we will recall some features of this
 group and argue that there is a magnetic counterpart to the sky group as well, which arises from the Bianchi identity. Together these two sets constitute the superselected operators of QED. 
 If we consider a compact region of $M^{N-1}$, say, a ball of finite radius, with the boundary $S^{N-2}$, these
 operators may also be thought of as generating the `edge states' of the theory.
 
All these structures also exist for the case of non-Abelian gauge theories such as QCD, with subtle consequences which are still not fully understood \cite{Balachandran:2019bzg}. Unlike the case of the gauge group ${\cal G}(U(1))$ of QED, the gauge group ${\cal G}(SU(3))$ of, say, QCD, is non-Abelian. Considerations of locality show that, in this case also, the observables must commute with it. As in QED, one may relate these operators to the edge states as well.
But there are additional consequences due to the non-abelian nature of the groups $\G(G)$.
We will see that superselection sectors will be labelled, at least partially, from a maximal abelian algebra or a complete commuting set (CCS) taken from $\mathbb{C}{\cal G}(G)$, the group algebra of ${\cal G}(G)$. 

The algebra of local observables $\A$
acting on a vector state labelled by the eigenvalues of a CCS, cannot change these eigenvalues. (Additionally, the state is left invariant by
the Casimir operators and the Cartan subalgebra of 
$\mathbb{C}{\cal G}(G)$ containing the chosen CCS.)
But a non-trivial representation of the generators of $G$ in ${\cal G}(G)$ do change these eigenvalues. So they cannot be implemented in an irreducible representation of ${\cal A}$, a feature similar to the spontaneous breaking of the Lorentz transformations in the charged sectors of QED.
 Notice that what is `broken' is the global group, such as $SU(3)$.
 
 As in the Abelian case, there is a magnetic analogue of $\mathbb{C}{\cal G}(G)$ or rather its Lie algebra 
$\mathbb{C}{\cal Q}(G)$, again following from the Bianchi identity, which we 
shall refer to as $\mathbb{C}{\cal M}(G)$. Its elements measure moments of the magnetic field at infinity so that the full superselection group becomes a Drinfel'd double. In section 3, we will formulate and explore features of this
generalized sky group.
It is interesting to ask how various features of
the `breaking' of $G$ in an irreducible representation of
$\A$ generalizes to the larger context, including the magnetic analogue
$\mathbb{C}{\cal M}(G)$. This is discussed in section 4.

The addition of the Chern-Simons term to the $(2+1)$-dimensional
gauge theory brings in a new feature: the group of ``large" gauge transformations gets centrally extended both for $G=U(1)$ or for a non-abelian $G$. 
This has been shown for the theory on a disk $D^2 \times$ time \cite{Balachandran:1993tm}.
A similar analysis works here. We discuss the Chern-Simons
twist in section 4 as well.
Because of the extended algebra,
Chern-Simons modified gauge theories have to be
dealt with as non-abelian theories with all the attendant subtleties. 

Another immediate and natural question which 
can arise is the following: Given $\mathbb{C}{\cal G}(G)$ and 
$\mathbb{C}{\cal M}(G)$, can we identify duality transformations connecting these, and if so, is there a ``magnetic description" which exchanges the role
of these?
In section 5, we construct a duality-invariant action
which realizes the algebra of $\mathbb{C}{\cal G}(G)$ and 
$\mathbb{C}{\cal M}(G)$ in a canonical framework. We show that
the standard
``electric description" and its dual, the ``magnetic description", can be obtained from this as special gauge choices.

As mentioned earlier, the algebra $\mathbb{C}Q(G) \vee \mathbb{C}\M(G)$
generates the edge states of the theory for a compact region. The definition
and meaning of local charges in this case are briefly discussed in section 6.
We have also mentioned that only a CCS belonging to $\mathbb{C}{\cal G}(G)$ can be unitarily implemented. If this is the case, what do we mean by, or how do we define, the notion of color confinement?
We discuss this in section 6 as well, arguing that the only meaningful definition is
that confinement is equivalent to the statement that the colored states do not belong to the domain of the Hamiltonian.

There are also certain additional structures we can identify within this framework, such as Cuntz algebras and Drinfel'd double.
We mention that
Cuntz algebras were introduced in algebraic quantum physics in the DHR papers \cite{Doplicher:1969tk,Doplicher:1969kp,Doplicher:1971wk,Doplicher:1973at} and are also discussed in  Doplicher and Roberts \cite{Doplicher:1990pn}. The latter used them to prove a deep generalization of Tanaka-Krein duality. 
The key point for us is that gauge
transformations can be localized. If we consider local observables ${\cal A}({\cal O})$ in a finite open causally complete region ${\cal O}$, such as a double diamond, we can localize all the gauge transformations to $\overline{\cal O}$ and consider $\overline{\cal O}$ as our spacetime, with $\partial{\cal O}$ as the substitute for the infinity of the standard spacetimes like $M^4$. Then for each such ${\cal O}$, there is a ``large" gauge group ${\cal G}(G; \partial {\cal O})$ which is just like ${\cal G}(G)$ on the full spacetime. There is also a 
magnetic analogue ${\cal M}(G;\partial{\cal O})$ which together encode a Cuntz algebra $O_d^0(G, \partial {\cal O})$. Once more using $\mathbb{C}{\cal G}(G; \partial {\cal O})$ and $O^0_d (G; \partial {\cal O})$, we see the emergence of both a Drinfel'd double and a Manin triple. An implication is that {\it each such local region ${\cal O}$  has edge states from these algebras and their representations}. As relativity has no role in this remark, it gives ``topological" excitations also in condensed matter systems.
These structures are discussed in the concluding section, although 
their full impact on the physics is still not clear. 

Superselection sectors are poorly understood in gauge theories. Besides the electric and magnetic charge superselection sectors, there are those due to infrared effects \cite{Frohlich:1978bf,Buchholz:1982ea} with theorems on the breaking of Lorentz symmetry in the charged sectors. There are also arguments that both the Lorentz group and the ``global symmetry" group $G$ are broken because of infrared effects when $G$ is non-abelian \cite{Balachandran:2015dva,Balachandran:2013wsa,Balachandran:2018jwf}. These effects are not expected in local field theories with only global symmetries as they require infrared photons or gluons. In contrast, both the local and global algebras of gauge theories share the superselection sectors from ${\cal G}(G)$ and ${\cal M}(G)$.
A more comprehensive understanding of superselection sectors is
clearly important, particularly since such sectors are also
fundamentally important for the epistomology of quantum 
field theory \cite{{Fioroni:1994kg},{Wightman:1995kq}}.

As argued beautifully by Fioroni and Immirzi \cite{Fioroni:1994kg} and Wightman \cite{Wightman:1995kq}, superselection sectors are fundamentally important for the epistemology of quantum physics. Thus measurements necessarily observe abelian subalgebras associated with ${\cal A}$ (see for instance \cite{landsmanbook} and references therein)  and these can be the algebras generated by the CCS labelling the superselection sectors. A better insight into their properties is thus called for.

\section{Asymptotic Charges and Magnetic Moments: Abelian Gauge Theories}

We start with a review of old and familiar material on the Gauss law when the gauge group ${\cal G}(U(1))$ is abelian and is associated with the global group $G=U(1)$.  
Our study of gauge transformations is in the canonical formalism, with the
splitting of Minkowski space $M^N$ as
$M^N = M^{N-1} \times \mathbb{R}$. On the
spatial slice $M^{N-1}$, we have the gauge potential
$A_i$ and the electric field $E^j$, which form a canonically conjugate pair.
 In addition, there are matter fields with charge density $J^0$. The Gauss law
\begin{equation}
\partial_i E^i + J^0 \approx 0
\label{gausslaw1}
\end{equation}
is a first class constraint in the classical theory. The analogous statement
in the quantum theory is that the elements of the algebra of observables ${\cal A}$ must commute with the Gauss law operators, that is to say they must be gauge-invariant.

A state $\omega$ on ${\cal A}$ is, as usual, a positive linear functional which is normalized to unity on the identity element $\mathbb{1} \in {\cal A}$,
i.e.,  $\omega(\mathbb{1}) = 1$. The states of interest to us are given by density matrices $\rho$, which are positive trace class operators with $\text{Tr } \rho =1$. (There may be more general states on ${\cal A}$, but the requirement of subadditivity excludes them.) When $\rho$ is of rank 1, we get a vector state, $\rho = |\psi \rangle \langle \psi |$, where $|\psi \rangle$ is a vector in the Hilbert space ${\cal H}$ on which ${\cal A}$ is represented. 
%A mixed state is a convex combination of pure states. 

It is worth remarking that while is often assumed that the Gauss law operator should vanish on the vectors $|\psi \rangle$, this is rather too restrictive.
For the purpose of maintaining gauge invariance, it is sufficient if
the action of the unitary transformation generated by the Gauss law operator acts as identity on $\ket{\psi}$ for `small' gauge transformations, 
and generate the superselection sectors for `large' gauge transformations
as we explain below.

A very simple example from quantum mechanics illustrates our point. Consider a $q$-bit with its observables $M_2(\mathbb{C})$ of two-by-two matrices. The $2\pi$-rotation $e^{i 2 \pi S_3}$ acting on an observable $m \in M_2(\mathbb{C})$ cannot change it so that we require $e^{i 2 \pi S_3} m e^{-i 2 \pi S_3} = m$ for all $m \in M_2(\mathbb{C})$. So $e^{i 2 \pi S_3} \in {\cal C}(M_2(\mathbb{C})$, the centre of $M_2(\mathbb{C})$. The centre consists of all diagonal entries with equal entries. Imposing also unitarity, we see that $e^{i 2 \pi S_3} = e^{i \theta} \mathbb{1}$. For a $q$-bit, $\theta = \pi$ is picked out so that $e^{i 2 \pi S_3} = -\mathbb{1}$. 

$e^{i 2\pi S_3}$ is a superselection operator, its values identifying superselection sectors. In states where $e^{i 2\pi S_3} \neq \pm 1$, one has anyons.

How does one measure $e^{i 2 \pi S_3}$ and get -1? For a pure state $|\psi \rangle \langle \psi|$, we can compute its mean value to get $\text{Tr }(|\psi \rangle \langle \psi | e^{i 2 \pi S_3})= \langle \psi |e^{i 2 \pi S_3}|\psi \rangle =-1$ as we want. This argument extends to all $q$-bit states.

Now a reasonably careful treatment of the Gauss law we outline below shows the existence of ``large" gauge transformations and their magnetic flux analogues which commute with ${\cal A}$. They generate the ``sky" group \cite{Balachandran:2013wsa} and its magnetic analogue and commute with ${\cal A}$ as well. But they need not vanish on vectors $|\psi \rangle$ which generate the density matrices $|\psi \rangle \langle \psi |$. Using the GNS construction, we can find a representation of ${\cal A}$ and of the sky group and its magnetic analogue as well. 

Turning to a more careful treatment of the Gauss law (\ref{gausslaw1}),
we introduce a test  function $\Lambda$ and the smeared operator
\begin{equation}
g(\Lambda) := -\int d^{N-1}x\, (\partial_i \Lambda) E^i (x) + \int d^{N-1}x \Lambda(x) J^0(x) \approx 0.
\label{gausslaw2}
\end{equation}
This is in accordance with the fact that derivatives of distributions are to be understood in terms of derivatives of test functions such as $\Lambda$.
But we would also need this constraint (\ref{gausslaw2}) to be consistent with the classical implementation of the Gauss law as in (\ref{gausslaw1}), so we require that the test functions $\Lambda$ should vanish at spatial infinity. 
This will ensure the passage from (\ref{gausslaw2}) to (\ref{gausslaw1}) via
an integration by parts.
So (\ref{gausslaw2}) is interpreted to hold for all $\Lambda \in C^\infty_0(\mathbb{R}^{N-1})$, the superscript $\infty$ indicating infinite differentiability, and the subscript $0$ indicating that a $\Lambda \in C^\infty_0$ vanishes at infinity. Thus
\begin{equation}
g(\Lambda) \approx 0 \quad \text{for all} \quad \Lambda \in C^\infty_0(\mathbb{R}^{N-1}).
\end{equation}

The smeared operator $g(\Lambda )$ is well defined even for test functions which are nonzero at spatial infinity, even though it does not vanish
as a constraint on observables.
We will denote test functions which do not necessarily vanish at
spatial infinity as $\Theta$, and define the corresponding
operators from (\ref{gausslaw2}) as $Q(\Theta)$; i.e.,
\begin{equation}
Q(\Theta)= \int d^{N-1}x \,\left[ -(\partial_i \Theta) E^i (x) + \Theta (x) J^0(x)\right], \hskip .2in \Theta(x) \in C^\infty(\mathbb{R}^{N-1}).
\label{charge}
\end{equation}
Our aim is to analyze the properties of this operator and a magnetic dual version of it,
which we shall introduce shortly.

First of all, notice that since
\begin{equation}
[g(\Lambda), Q(\Theta)] = 0,
\end{equation}
$Q(\Theta)$ qualifies as an element of the set of operators of interest.
Further, if $\Theta$ vanishes at infinity, i.e., if $\Theta \in C^\infty_0(M^{N-1})$, then $Q(\Theta)$
is identical to $g(\Lambda)$ and it vanishes on all vector states. For test functions $\Theta$ which go to a constant $\Theta^\infty$ at infinity, $Q(\Theta )$ is proportional to the electric charge $Q(1)$, $Q(\Theta) = \Theta^\infty Q(1)$. If it goes to an angle-dependent function at infinity,
\begin{equation}
\Theta(\vec{x} = r \hat{x}) \rightarrow \Theta^\infty(\hat{x}),
\end{equation}
we get the elements of what was referred to as the sky group in \cite{Balachandran:2013wsa}.

Another important property of $Q(\Theta)$ is that it commutes with all local observables. Let $\phi(x)$ be a local tensorial field and $f$ a test function supported in ${\cal O}$. Then $\phi(f) = \int d^{N-1}x \bar{f}(x) \phi(x)$ is localized in a bounded open region ${\cal O}$ with compact closure $\overline{\cal O}$, and $\Lambda$ is the restriction of $\Theta$ to $\overline{\cal O}$, then by the locality of commutators,
\begin{equation}
[Q(\Theta), \phi(f)] = [Q(\Lambda), \phi(f)] .
\end{equation}
But now $Q(\Lambda) = g(\Lambda)$ so that
\begin{equation}
[Q(\Theta), \phi(f)]=0.
\end{equation}
In other words, $Q(\Theta)$ commutes with local observables. 

The set of operators  $\{e^{i Q (\Theta )}\}$ form the quotient of the gauge algebra
by `small' gauge transformations (which correspond to
$e^{i g (\Lambda)}$, with $\Lambda \in C^\infty_0$).
Since $Q(\Theta )$ commute with local observables, the full set of
observables is made of a complete commuting set from 
$\{ e^{i Q (\Theta )}\}$ and the local observables $\A$.
The complete commuting set from $\{ e^{i Q (\Theta )}\}$ will define the superselection sectors of the theory. In the present case of an Abelian gauge theory, the entire set $\{ e^{i Q (\Theta )}\}$, for all
$\Theta$, form the complete commuting set. Thus ${\cal A}$, which includes $e^{iQ(\Theta)}$ in its centre ${\cal C}({\cal A})$ for a fixed ${\cal A}$, is not a factor \footnote{Although we use the same symbol ${\cal A}$ for the local algebra and the local algebra extended by $\{e^{iQ(\Theta)}\}$, the meaning should be clear from the context.}

There is an elegant way to write (\ref{charge}). In four dimensions, (i.e., for
$N =4$), the Maxwell equation corresponding to the Gauss law, written
in terms of differential forms, reads
\begin{equation}
d\left[ {\textstyle{1\over 2}}\, {}^*F_{ij} dx^i \wedge dx^j \right] \equiv d{}^*F = J^0\, d^3 x .
\end{equation}
So we can write (\ref{charge}) as
\begin{equation}
Q(\Theta) = \int_{\mathbb{R}^3} d (\Theta^*F).
\label{elecQ}
\end{equation}
More generally, for arbitrary number of dimensions, we can write this as
\begin{equation}
Q(\Theta) = \int_{\mathbb{R}^{N-1}} d (\Theta^*F).
\end{equation}

In four dimensions, we can clearly write a dual version of 
(\ref{elecQ}) defined as
\begin{equation}
M(\Phi) = \int_{\mathbb{R}^{3}} d (\Phi F)
= \int_{\mathbb{R}^{3}} d \Phi \wedge F
\end{equation}
with a test function $\Phi \in C^\infty(\mathbb{R}^{3})$. 
The second equality follows from the Bianchi identity.
For those functions $\Phi$ which vanish at spatial infinity,
evidently,
\begin{equation}
M(\mu) \approx 0.
\end{equation}
 We have used
$\mu$ to denote test functions which vanish at 
infinity, i.e., $\mu \in C^\infty_0(\mathbb{R}^{3})$.
The operators $M(\Phi)$ need not be weakly zero for
$\Phi \in C^\infty(\mathbb{R}^{3})$, which do not vanish at infinity.
With such test functions, $M(\Phi)$ generate an abelian subalgebra, which is "localized at infinity", since formally
\begin{equation}
M(\Phi) = \int_{\partial \mathbb{R}^{3}} \Phi F.
\label{mcharge}
\end{equation}
As a result, $M(\Phi )$ commutes with all local observables.
Thus they form another set of superselected operators
for the theory.
So $M(\Phi) \in {\cal C}({\cal A})$, the centre of the algebra of observables.

In this way, for a $U(1)$ gauge theory, we obtain an infinite number of superselected operators, $Q(\Theta)$ and $M(\Phi)$, 
which are characterized by the asymptotic limits of $\Theta$ and $\Phi$.

Also, from canonical commutation relations, we can work out
the commutator algebra of the superselected operators 
$\{Q(\Theta), M(\Phi)\}$. It is given by
\begin{equation}
[Q(\Theta_1), Q(\Theta_1)] = [Q(\Theta_1), M(\Phi)] = [M(\Phi_1), M(\Phi_2)]=0.
\label{supalg1}
\end{equation}

To summarize the key results of this section, we have two sets of superselected operators, $Q(\Theta)$ and $M(\Phi )$, which obey the
algebra (\ref{supalg1}) given above. 
Clearly the interesting question is about the observable consequences of 
these operators. We will come back to this after discussing the nonabelian case, which also possesses a similar set of superselected operators.

\section{Asymptotic Charges and Magnetic Moments: Non-abelian Gauge Theories}

We take the gauged ``symmetry" group $G$ to be a compact connected simple Lie group such as $SU(3)$, and the spacetime as before to be the Minkowski manifold $\mathbb{R}^N = \mathbb{R}^{N-1} \times \mathbb{R}^1$.
On the spatial slice $\mathbb{R}^{N-1}$, we have the Lie algebra valued
connection one-form $A_i dx^i$ and its canonically conjugate field $E^j$.
For specificity, we use a set of matrices
$t_a$ which span the Lie algebra 
$\underline{G}$ of $G$ in its defining representation (such as the triplet
representation for $SU(3)$) and fulfilling the normalization condition
\begin{equation}
\text{Tr } (t_a t_b ) = \half \delta_{ab}.
\label{NA1}
\end{equation}
The fields $A_i$ and $E^j$ can be expanded in this basis as
\begin{equation}
A_i = A_i^a t_a, \hskip .3in
E^i = E^{i,a} t_a .
\label{NA2}
\end{equation}
The nonabelian version of the Gauss law (\ref{gausslaw1}) is
\begin{equation}
D_i E^i + J^0 \approx 0
\label{gausslawnab1}
\end{equation}
where $J^0$ is the Lie algebra valued charge density from matter fields,
which can also be expanded in terms of $\{ t_a\}$ as $J^0 = J^{0,a} t_a$,
and $D_i$ denotes covariant differentiation,
\begin{equation}
D_i E^i := \partial_i E^i -i  [A_i, E^i].
\label{covder}
\end{equation}
As before, we rewrite (\ref{gausslawnab1}), after smearing it with test functions $\Lambda = \Lambda^a t_a$,  with $\Lambda^a \in C^\infty_0({\mathbb R}^{N-1})$, as
\begin{equation}
g(\Lambda) = \int d^{N-1} x\, \text{Tr } [-D_i \Lambda E^i + \Lambda J^0] \approx 0.
\label{gausslawnab2}
\end{equation}
Since $\Lambda$ vanish on $S^{N-2}$ as $r \rightarrow \infty$,
the commutator of $g (\Lambda )$ with any operator reduces to the commutator of the Gauss law with the same operator. Thus
all the observables must commute with $g(\Lambda)$. 

We can now write down the non-abelian charges by extending the left hand side of (\ref{gausslawnab2}) to test functions $\Theta = \Theta^a t_a$, where $\Theta^a \in C^\infty({\mathbb R}^{N-1})$ do not necessarily vanish at infinity. Explicitly, the charges are thus given as
\begin{equation}
Q(\Theta) = \int d^{N-1}x\, \text{Tr } (-D_i \Theta\, E^i + \Theta J^0).
\label{chargenab}
\end{equation}
$Q(\Theta)$ is not weakly zero if $\Theta^a \notin C^\infty_0({\mathbb R}^{N-1})$.

Turning to observables, as mentioned above,
the local observables commute weakly with $Q(\Theta)$. 
It is useful to work again through the reasoning for this result.
Recall that a local observable is either a field $\psi(x)$ at a point $x$, or its smeared version
\begin{equation}
\psi(f) = \int_{\cal O} d^{N-1} x\, f(x) \psi(x),
\end{equation}
where ${\cal O}$ denotes a bounded open region
which is the support of $f(x)$.
The locality of the commutation relations implies that the commutator of $Q(\Theta)$ with a local observable 
depends only on $\Theta|_{\cal O}$, the restriction of
$\Theta$ to $\O$.
 We can smoothly change $\Theta$ outside of ${\cal O}$ to zero, getting 
a new function $\hat{\Theta} \in C^\infty_0({\mathbb R}^{N-1})$.
Thus, effectively, $Q(\Theta)$ in the commutator with $\psi (f)$
can be taken to be $Q({\hat\Theta})$ and as a result,
\beqar
[Q(\Theta), \psi(f)] &=& [Q(\hat{\Theta}), \psi(f)] = [g(\hat{\Theta}), \psi(f)] 
\approx 0.
\label{NA4}
\eeqar
We see thus that local observables weakly commute with the 
charges $Q(\Theta)$ and that $Q(\Theta)$ are superselected operators for local observables. But the key point of distinction with the previous case is
that $Q({\Theta})$
generate {\it non-abelian} superselection rules since the commutator
\begin{equation}
[Q(\Theta_1), Q(\Theta_2)]  \approx Q([\Theta_1,\Theta_2])
\label{NA5}
\end{equation}
is not identically zero.
(Henceforth, we drop the symbol $\approx$ as we will deal with quantum theory.)

Using the Yang-Mills field equation
$D^*F = J^0$,
we can write (\ref{chargenab}) as
\begin{equation}
Q(\Theta) = \int d\, \text{Tr } (\Theta\, ^*F).
\label{NA7}
\end{equation}
Again, in four dimensions, this naturally leads to a dual set of operators
\beq
M(\Phi) = \int d\, \text{Tr } (\Phi F ), \label{mchargenab} 
\eeq
where $\Phi = \Phi^a t_a$, with $\Phi^a \in C^\infty({\mathbb R}^{3}) $.
If $\mu = \mu^a t_a$ vanishes at infinity, i.e., $\mu^a \in C^\infty_0({\mathbb R}^{3})$, then clearly $M(\mu ) = 0$.
Also, as before, using Bianchi identity $D F=0$, we can rewrite $M(\Phi)$ in four dimensions as
\begin{equation}
M(\Phi) = \int \text{Tr } (D \Phi) \wedge F.
\label{NA8}
\end{equation}
We can generalize this equation to dimension $N >4$,
by choosing $\Phi$ to be a differential form of rank $N-4$, but for most of 
the following discussions, we take $N= 4$.

The components $F^a_{ij}$ of the magnetic field $\half F_{ij} dx^i \wedge dx^j$ commute among themselves on the spatial slice $\mathbb{R}^{3}$, hence $M(\Phi)$'s generate an Abelian algebra; i.e, 
\begin{equation}
[M(\Phi_1), M(\Phi_2)]=0, \hskip .2in \Phi_1^a, \,
\Phi_2^a \in C^\infty({\mathbb R}^{3}).
\label{mmcomm}
\end{equation}
Also, as is evident from (\ref{mchargenab}), $M(\Phi)$ is a surface term at infinity. Therefore, it commutes with all local observables and form another set of superselected operators.

However, unlike the abelian case, $Q(\Theta)$ and $M(\Phi)$ need no longer commute; rather we obtain the commutation relation
\beq
[Q(\Theta), M(\Phi)] = M([\Theta, \Phi]),  \label{chargecomm1}
\eeq

Returning to the algebra of the $Q(\Theta)$, notice that we have the
algebraic relations
\beqar
\protect{[}Q(\Theta_1), Q(\Theta_2)] &=& Q([\Theta_1, \Theta_2]),  \label{chargecomm2} \\
\protect{[}Q(\Theta), g(\Lambda)] &=& g([\Theta, \Lambda]),  \quad\quad \text{since} ~[\Theta, \Lambda] \in C^\infty_0(M^{N-1}), \label{chargecomm3} \\
\protect{[}g(\Lambda_1), g(\Lambda_2)] &=& g([\Lambda_1, \Lambda_2]). \label{chargecomm4}  
\end{eqnarray}
From (\ref{chargecomm3}) above, we see that $e^{i g(\Lambda)}$ generate a normal subgroup of the group of transformations
$e^{i Q(\Theta)}$. In the representation of these operators on the Hilbert space, $g(\Lambda) \rightarrow 0$ and $e^{i g(\Lambda)} \rightarrow \mathbb{1}$. Hence the quotient $\{ e^{i Q(\Theta)} \} / \{ e^{i g(\Lambda)} \}$ defines operators which intuitively are ``localized at infinity". It is this group which is relevant in the quantum theory, since, by Gauss law, $g(\Lambda) \rightarrow 0$ on the Hilbert space.
We will refer to the transformations generated by
$Q(\Theta)$, $M(\Phi)$, with the commutation rules
(\ref{mmcomm}), (\ref{chargecomm1}), (\ref{chargecomm2}) as the
{\it generalized sky group}.

The exponentials $e^{i Q(\Theta)}$ give the group of maps from $\partial M^{N-1} = S^{N-2}$ to $G$ as $g(\Lambda)$ 
vanishes. From this fact and the algebraic relations
({\ref{mmcomm}), (\ref{chargecomm1}),  we conclude
that the set of superselection operators is given by the semi-direct product of Maps$(S^{N-2}, G)$ with the abelian algebra 
of $M(\Phi)$'s. Once again, the key distinction with the case of abelian
theories is that these generate a {\it non-abelian} superselection algebra.

Some further reduction will be useful before we can deduce some of the consequences of the
nonabelian nature of the superselection operators.
Towards this, let us set
\beq
\Theta^\infty ({\hat x})  = \lim_{r \rightarrow \infty} \Theta (r \hat{x}), \hskip .2in
\Phi^\infty ({\hat x})  = \lim_{r \rightarrow \infty} \Phi (r \hat{x}).
\label{NA9}
\eeq
Thus $\Theta^\infty$, $\Phi^\infty$ are (Lie algebra valued) functions
on the sphere $S^{N-2}$.
On the Hilbert space of states, by virtue of the Gauss law and the Bianchi identity,
$Q(\Theta)$ and $M(\Phi)$ depend only on $\Theta^\infty$ and $\Phi^\infty$, so that they give operators $\hat{Q}(\Theta^\infty), \hat{M}(\Phi^\infty)$; i.e.,
\beq
Q(\Theta) \rightarrow \hat{Q} (\Theta^\infty), \hskip .2in
M(\Phi) \rightarrow \hat{M}(\Phi^\infty).
\label{NA10}
\eeq
The commutation relations can also be written in terms of
$\Theta^\infty$, $\Phi^\infty$ as
\begin{eqnarray}
[\hat{Q}(\Theta^\infty_1), \hat{Q}(\Theta^\infty_2)] &=& \hat{Q}([\Theta^\infty_1,\Theta^\infty_2]) \nonumber\\
\protect{[}\hat{Q}(\Theta^\infty), \hat{M}(\Phi^\infty)] &=& \hat{M}([\Theta^\infty,\Phi^\infty]).
\label{NA11}
\end{eqnarray}
We see that we have an infinite-dimensional algebra of the superselection operators
labeled by Lie algebra valued functions on the sphere
$S^{N-2}$. For our analysis, it is important that the
nature of the commutation relations for this infinite-dimensional
superselection algebra
is entirely determined by the structure of the Lie algebra 
$\underline{G}$ of the finite-dimensional group $G$.
In fact, explicitly, we can write
\beq
\Theta^\infty_1 = \Theta^{\infty,a}_1 t_a,
 \quad \Theta^\infty_2 = \Theta^{\infty,a}_2 t_a,\quad
\Phi^\infty = \Phi^{\infty,a} t_a,
\label{NA12}
\eeq
where $
 \Theta^{\infty,a}_1$, $\Theta^{\infty,a}_2$, $\Phi^{\infty,a}$
 are elements of  $\text{Maps}(S^{N-2}, \mathbb{C})$.
 (We can complexify the functions as well, as indicated by the target space
 of these maps.)
The test functions for the right hand side
of (\ref{NA11}) then become
\begin{eqnarray}
[\Theta^\infty_1, \Theta^\infty_2] &=& \Theta^{\infty,a}_1 \Theta^{\infty,b}_2 [t_a, t_b] \label{asym1}\\
\protect{[}\Theta^\infty, \Phi^\infty] &=& \Theta^{\infty,a} \Phi^{\infty,b} [t_a, t_b] \label{asym2}
\end{eqnarray}
showing explicitly how the Lie algebra commutation rules determine the structure of the superselection algebra (\ref{NA11}).

We can now define a maximal abelian subalgebra from $\mathbb{C} \text{Maps} (S^{N-2}, G^{\mathbb{C}})$. If $H_i$ span a Cartan subalgebra of $\underline{G}$, then from (\ref{asym1},\ref{asym2}), we see that $\Theta^{\infty,j}_1 H_j, \Theta^{\infty,j}_2 H_i$ and $\Phi^{\infty,i} H_i$ commute and, hence,
$\hat{Q}(\Theta^{\infty,j}_\alpha H_j)$, $\hat{M}(\Phi^{\infty,i} H_i)$
generate an abelian algebra. It is infinite-dimensional as the functions $\Theta^{\infty,j},\Phi^{\infty,j}$ span an infinite-dimensional space, being maps from $S^{N-2}$ to $\mathbb{C}$. We can take these functions to
belong to a Hilbert space $L^2(S^{N-2})$ with Lie algebra-valued elements by introducing the scalar product
\begin{equation}
(\Theta^{\infty,j}_1,\Theta^{\infty,j}_2) = \int_{S^{N-2}} d \Omega_{\hat{n}} \,\text{Tr } \bigl(\Theta^{\infty,j}_1 (\hat{n})^* \Theta^{\infty,j}_2(\hat{n})\bigr), \quad \hat{n} \in S^{N-2} 
\label{NA13}
\end{equation}
where $d\Omega_{\hat n}$ is the rotationally invariant volume form on $S^{N-2}$. 
It is convenient to choose the normalization
\begin{equation}
\int_{S^{N-2}} d \Omega_{\hat{n}} =1.
\label{NA14}
\end{equation} 

Let $\hat{\Theta}^\infty_n$ define an orthonormal basis on $S^{N-2}$ for $L^2(S^{N-2})$, where $n$ now denotes a generic index for the
basis elements. We write
\begin{equation}
\hat{Q}(\hat{\Theta}^\infty_n) = \hat{Q}_n, \quad \hat{M} (\hat{\Theta}^\infty_n) = \hat{M}_n.
\label{NA15}
\end{equation}
The operators $\{\hat{Q}_n, \hat{M}_n \}$
generate an abelian algebra.

It is convenient to choose the basis functions $\hat{\Theta}^\infty_n$ to be real, with $\hat{\Theta}^\infty_0$ as the constant function, equal to 1,
on $S^{N-2}$, whose integral over $S^{N-2}$ is also 1, since we have the
normalization (\ref{NA14}).
With these choices, we also get
\begin{equation}
\hat{Q}_n^* = \hat{Q}_n, \quad \hat{M}_n^* = \hat{M}_n.
\label{NA16}
\end{equation}

\section{Realizations of the superselection algebra}

We now turn to some of the consequences of the nonabelian nature of the superselection algebra. Some of the results to follow were anticipated in
\cite{Balachandran:2013wsa}.

\subsection{${\cal G}(G)$ is spontaneously broken}

From the arguments in the previous section, we see that a vector state in a superselection sector can be labeled by the eigenvalues $\hat{q}_n, \hat{m}_m$ of a complete commuting set (CCS) such as $\hat{Q}_n,\,\hat{M}_m$, with $m, n \in \mathbb{Z}$, and possible additional operators. Any element  in the local algebra ${\cal A}$ acting on such a vector state cannot change the eigenvalues $\hat{q}_n, \hat{m}_m$ since it commutes with $\hat{Q}_n,\hat{M}_m$. 
Thus $\{ \hat{q}_n, \hat{m}_m\}$ are labels for a superselection sector 
of the local algebra ${\cal A}$.

But the group ${\cal G}(G)$ has operators which do not commute with $\hat{Q}_n, \hat{M}_m$. If $E_\alpha, E_{-\alpha}$ are the roots of the Lie algebra $\underline{G}$ of $G$, it is enough to consider test functions
$\Theta^\pm = \Theta^{\pm \alpha} E_{\pm \alpha}$
and the corresponding charges and magnetic fluxes 
\begin{equation}
Q(\Theta^{\pm \alpha} E_{\pm \alpha}) = Q^\pm, \quad M(\Theta^{\pm \alpha} E_{\pm \alpha}) = M^\pm.
\label{NA17}
\end{equation}
Clearly, these {\it do not} commute with the chosen set
$\{ {\hat Q}_n, {\hat M}_m \}$ and so change the eigenvalues labeling a superselection sector. Thus $\{ Q^\pm, M^\pm \}$
  {\it cannot} be implemented as operators in a given superselection sector of ${\cal A}$.
  In other words, the group ${\cal G}(G)$ is {\it spontaneously broken} to the group generated by $\hat{Q}_n,\hat{M}_m$. We will denote the unbroken group as 
  ${\cal G}(\hat{Q}_n,\hat{M}_m)$.

This breakdown of $\mathbb{C}{\cal G}(G)$, which includes the representatives of the global group $G$, to one of its maximal abelian subalgebras, say $\mathbb{C}{\cal G}(\hat{Q}_n,\hat{M}_m)$, is, at one level, similar to the Higgs field $\phi$ breaking a symmetry group $G$ to the maximal stability group $G_c$ of its vacuum expectation value $\langle \phi \rangle$. The group $G_c$ is also the stability group of the asymptotic value $\phi^\infty$ of $\phi$, so that we see that $G_c$ also labels the superselection sectors of the associated local observables. 

But there are also differences. The unbroken algebra in the gauge theory case is always abelian. But for symmetry breaking via the
Higgs field, $G_c$ can be non-abelian. An example is where $G=SU(3)$, 
$\phi^\infty=$ constant$\times Y$ ($Y= {2 t_8/\sqrt{3}}$ is the hypercharge) and $G_c = U(2) =(SU(2) \times U(1))/\mathbb{Z}_2 \subset G$. 
Again for the standard model, $G = [SU(3) \times SU(2) \times U(1)]/\mathbb{Z}_6$ and ${\cal G}(\hat{Q}_n,\hat{M}_m)$, uses the commuting generators $I_3$, $Y$ of $SU(3)$, $I_3$ of $SU(2)$ and the weak hypercharge, whereas the Higgs field breaks it to $U(3)$.
Further, there are no Goldstone modes or gauge boson mass generation
for the symmetry breakdown $\mathbb{C}{\cal G}(G) \rightarrow
\mathbb{C}{\cal G}(\hat{Q}_n,\hat{M}_m)$.

There is a deficiency in this analysis. We need a state $\omega$, a positive linear functional on ${\cal A}$ with $\omega(\mathbb{1}) =1$, such that the expectation values $\omega(e^{i Q(\Theta)})$ and $\omega(e^{i M(\Phi)})$ are not zero when $\Theta^\infty$ and $\Phi^\infty$ are not zero. We plan to take up this issue in detail in another paper. But see the next subsection for remarks in this direction.

\subsection{The Chern-Simons Twist in $3+1$ Dimensions}

Witten considered this case in his paper \cite{Witten:1979ey} showing that magnetic monopoles acquire fractional charge proportional to $\theta$ in the presence of $\theta$-vacua (This $\theta$ is not our test function). We will formulate his approach in another way.

Thus let $G$ be our global group, $A$ its connection field and $|0\rangle$ its vacuum vector state. Consider the twisted state
\begin{equation}
|\theta \rangle = \exp \left(i \theta \int \text{Tr } (A dA + \frac{2}{3} A^3)\right) |0\rangle
\label{NA18}
\end{equation}
(In this expression, $A_i dx^i  = - i t_a A^a_i dx^i$.)
Since the `small' gauge transformations generated by $g(\Lambda)$ do not change $|\theta \rangle$, we have that $g(\Lambda) |\theta \rangle =0$. Now our ${\cal A}|\theta \rangle$ gives us a vector space giving a representation 
of ${\cal A}$. We assume that the scalar product of these vectors can be calculated by quantum field theory techniques; the resultant space can be completed into a Hilbert space.

Consider next the action of $e^{i Q(\Phi)}$ on these states. Using (\ref{NA11}), one gets just a boundary term 
\begin{equation}
e^{i Q(\Phi)}{\cal A}\,|\theta \rangle = e^{ i \theta M(\Phi)}{\cal A}\,|\theta \rangle.
\label{qontheta}
\end{equation}
Thus if the magnetic flux at infinity is non-zero, the eigenvalue of $Q(\Phi)$ is $\theta M(\Phi) \mod 2\pi$.

In the presence of the 't Hooft-Polyakov monopole, where the global $G$ is broken to $U(1)$, $M(\Phi)$ is, in fact, not zero. Classically $F$ is the magnetic field of the monopole and in $M(\Phi)$, we take the moments of this field on $S^2$ from which we can reconstruct (the asymptotic form of ) $F$ itself. For a spherically symmetric monopole with a Higgs triplet, the connection one-form on $S^2$ fulfills
\begin{equation}
D( \vec{\tau} \cdot \hat{n}) = d (\vec{\tau} \cdot \hat{n}) + [A, \vec{\tau} \cdot \hat{n}] = 0,
\label{NA19}
\end{equation}
so that, with $\tau_i$ as the Pauli matrices,
\begin{equation}
A = \frac{1}{2} (\vec{\tau} \cdot \hat{n}) d (\vec{\tau} \cdot \hat{n}), \quad F = -\frac{1}{4} \vec{\tau} \cdot (d \hat{n} \wedge d \hat{n}).
\label{AandF}
\end{equation}
In the quantum theory, $M(\Phi)$ is an operator and we need a state where the expectation values lead to (\ref{AandF}). 
%We do not know what exactly such a state is. 

But semiclassically, we can interpret $M(\Phi)$ as the eigenvalue of $Q(\Phi)$. In that case, the unbroken $U(1)$ subgroup has the generator $\tau \cdot \hat{n}$ and the charge is
\begin{eqnarray}
\theta M(\vec{\tau} \cdot \hat{n}) &=& -\frac{\theta}{4} \int_{S^2} \text{Tr } (\vec{\tau} \cdot \hat{n}) \vec{\tau} \cdot (d \hat{n} 
\wedge d \hat{n}) \\
&=& -\frac{\theta}{2} \int_{S^2} \hat{n} \cdot (d \hat{n} \wedge d \hat{n}) = -2\pi \theta,
\end{eqnarray}
which is Witten's result.

\subsection{The Chern-Simons Twist in $2+1$ Dimensions}

In this case, one adds the Chern-Simons term to the Lagrangian. It has been treated by us in a few papers \cite{Balachandran:1991dw,Balachandran:1992yh} previously when the spatial slice is a disk $D_2$ of radius $R$. In that case, the gauge algebra is localized on $\partial D_2 =S^1$ and fulfils the Kac-moody algebra
\begin{equation}
[Q(\Theta_1), Q(\Theta_2] = Q([\Theta_1,\Theta_2]) + k \int_{S^1} \text{Tr }( \Theta_1 d \Theta_2).
\label{kmalg}
\end{equation}
So $Q(\Theta)$ are generators of the Kac-Moody algebra with the central term of level $k$ if the Chern-Simons term in the Lagrangian is
\begin{equation}
S_{\rm C.S.} = k \int \text{Tr } (A d A + \frac{2}{3} A^3).
\label{csaction}
\end{equation}
Since (\ref{csaction}) involves no term with the scale $R$, we can take its limit as $R \rightarrow \infty$. So the edge algebra for $Q(\Theta)$ is also 
(\ref{kmalg}), in the limit of $R \rightarrow \infty$.
It is remarkable that even for $G=U(1)$, we get a nonabelian gauge algebra of charges.

For $U(1)$, $Q(\hat{\Theta}^\infty)$ generates an abelian algebra. For $\hat{\Theta}^\infty =$ constant, it is in the centre of the full Kac-Moody algebra. Since $Q(\hat{\Theta}^\infty) = \hat{\Theta}^\infty Q(\mathbb{1})$ and $Q(\mathbb{1})$ is the charge, we see that charge acts as a label for the abelian Kac-Moody irreducible representation. This is a basic and known result, but is derived here in the present approach.

If $G$ is non-abelian, we get 
\begin{equation}
[\hat{Q}(\hat{\Theta}^\infty_1), \hat{Q}(\hat{\Theta}^\infty_2)] = \hat{Q}([\hat{\Theta}^\infty_1, \hat{\Theta}^\infty_2]) + k \int_{S^1} \text{Tr }(\hat{\Theta}^\infty_1 d \hat{\Theta}^\infty_2)
\label{kmalgnab}
\end{equation}
if the Chern-Simons term in the Lagrangian is as in (\ref{csaction}). So again, the charges, for which 
$\hat{\Theta}^\infty$ are constant, do not have the $k$-term and $Q(t_a)$ generate the Lie algebra of $G$.
We note that (\ref{kmalgnab}) is the standard form of the Kac-Moody Lie algebra. The highest weight representations of this algebra are reviewed in \cite{Goddard:1986bp}.

The central term commutes with $M(\Phi)$ and does not affect the commutators between $Q(\Theta)$ and $M(\Phi)$, for example,
 in verifying the Jacobi identity,
\begin{equation}
[[\hat{Q}(\hat{\Theta}^\infty_1), \hat{Q}(\hat{\Theta}^\infty_2)], \hat{M}(\Phi^\infty)] +\text{cyclic} =0.
\end{equation}

\section{A duality-invariant edge action}
In this section we will make some further observations
on the electric-magnetic type duality which relates
 the operators
$Q(\Theta)$ and $M(\Phi)$.
A natural and interesting question regarding these operators
is whether we can construct an action
which has manifest duality-invariance and which leads to the
previously obtained algebra. The most interesting situation will be,
of course, for the
four-dimensional theory; this is the case we shall focus on. 

The quest for actions which display manifest duality-invariance has a long history, going back to the work
of Schwinger and Zwanziger in QED \cite{SZ}. The evaluation of QED partition
functions in a dual-symmetric way was also analyzed many years ago
by Witten \cite{wit-dual} and by Verlinde \cite{Ver}. Extensions to the
${\cal N} = 4$ supersymmetric Yang-Mills theory have also been
considered in \cite{Dor}.
Nevertheless efforts towards the construction of an action with manifest
duality-symmetry have not really been successful, although we may note
that, for the abelian theory, it is possible to obtain such an action by
dimensional reduction from a six-dimensional theory \cite{{Ver},{GN}}.
Here we would like to pose a more limited question: If we are only interested in the edge modes, i.e., the algebra of $Q(\Theta)$ and $M(\Phi )$,
is it possible to have an action with manifest duality symmetry?
We will see that the answer to this question is in the affirmative,
it is the subject of this section.

We start with the statement that the standard 
YM action (perhaps with the addition of the $\theta$-term, for full generality)
is sufficient to give a complete characterization of the physics we are interested in.
Thus, even though the theory is presented in what might be called the
``electric description", it should be fully adequate to describe 
the dynamics.
Thus any attempt to make a duality-invariant theory must 
take account of this fact.
This means that electric and magnetic descriptions must appear as equivalent
but different choices; i.e., they must be like different choices for
some larger gauge-like symmetry. We will construct an action which
has this property.

Let us start with the Yang-Mills action and choose the $A_0 = 0 $ gauge.
The spatial components of the gauge potential may be
parametrized as
\beq
A_i = U a_i U^{-1} - \del_i U \, U^{-1}
\label{8.1}
\eeq
with $U(x) \in G$.
This is not simply a gauge transformation of $a_i$, since
$U$ will be taken to be time-dependent in general. The components
$a_i$ must obey one constraint such as
of transversality ($\del_i a_i = 0$) or something similar to it, to
avoid redundancy of field variables.
With such a restriction on $a_i$, (\ref{8.1}) is a general parametrization for the gauge field. 

Using this parametrization and the standard Yang-Mills
action, we can identify the canonical one-form $W$ of the theory as
\beqar
W &=& - {2\over e^2} \int d^3x \Tr [ \delta A_i \, F_{0i} ]\nonumber\\
&=&  {2\over e^2}\int \left[  \Tr [ D_i (\delta U U^{-1}) F_{0i} ] 
+ \Tr [ U \delta a_i U^{-1} \, F_{0i} ] \right]\nonumber\\
&=&
- {2\over e^2}\left[  \int_{S^2} d \Omega \,\Tr [ \delta U U^{-1} F_{0i} n^i r^2 ]
+ \int \Tr [ U \delta a_i U^{-1} \, F_{0i} ]  + {\rm Gauss~law}\right].
\label{8.2}
\eeqar

We are interested in edge states, which are generated by the 
charge rotations $U$. This means that, on the space of fields
$\{ U, a_i, F_{0i} \}$, we can restrict to the subspace where
$\delta a_i =0$, with a suitable restriction on its conjugate variable.
Notice that the magnetic field, according to (\ref{8.1}), has the form
\beq
F =  d A = U \left( d a + a\,a \right) U^{-1}.
\label{8.2a}
\eeq
Asymptotically, the magnetic field in any monopole sector has the
form $F = U F^D U^{-1}$, where 
$F^D$, which is defined by $a_i$, corresponds to configurations
of the Dirac monopole type in
any set of chosen $U(1)$ subgroups of $G$, as in the GNO ansatz
\cite{GNO}.  Thus considering fixed
values for $a_i$ is equivalent to considering a fixed configuration
asymptotically, apart from an overall charge rotation.
This is the case we are considering. The canonical one-form, for action on states which obey the Gauss law constraint, may then be taken as
\beq
W = - {2\over e^2} \int_{S^2} d \Omega \,\Tr [ \delta U U^{-1} F_{0i} n^i r^2 ].
\label{8.2b}
\eeq
Define $F_{0i} n^i r^2 = E$. Further, we define the complex parameter
\beq
\tau = {\theta \over 2\pi} + i\, {4 \pi \over e^2} .
\label{8.3}
\eeq
We can then write
\beq
W = {i \over e^2 {\rm Im} \tau} \int \Tr [ \delta U U^{-1} \, E (\tau - {\bar\tau})].
\label{8.4}
\eeq
The addition of the $\theta$-term gives the one-form
\beq
W = {i \over e^2 {\rm Im} \tau}  \int \Tr \left[  \delta U U^{-1} \,\Bigl[  E (\tau - {\bar\tau})
+ i B (\tau + {\bar\tau}) \Bigr]
\right].
\label{8.5}
\eeq
The term $B (\tau + {\bar \tau}) = \theta B /\pi$ arises from the
$\theta F {\tilde F}$ term in the action.

Even though we obtained this from the YM action plus the $\theta$-term, we can now write an action for fields defined
on the boundary $S^2$ which leads to this canonical structure.
It is given by
\beq
S = {i \over e^2 {\rm Im} \tau} \int dt \int  \Tr \left[  \del_0 U U^{-1} \,\Bigl[  E (\tau - {\bar\tau})
+ i B (\tau + {\bar\tau}) \Bigr]
\right].
\label{8.6}
\eeq
It is trivial to check that the
canonical one-form obtained from (\ref{8.6}) agrees with
(\ref{8.5}).
As we will see again shortly, the electric field-dependent term in (\ref{8.6})
is also the action which comes from exponentiating the action of
$Q(\Theta)$ on the wave functions.
In this sense, the theory is indeed the standard YM theory
with a $\theta$-term.
Our goal now is to
construct a duality-invariant generalization of this.

As mentioned earlier, the ``electric description" given by
(\ref{8.6}) must emerge as one gauge choice (within a larger gauge symmetry)
of a duality-invariant action. This calls for an enlargement of
the space of fields.
So, analogous to $U$ we introduce a group-valued field
$V$ and define
\beqar
I_0 &=& D_0 U \, U^{-1} = \del_0 U \, U^{-1}  + U \, \alpha_0 \, U^{-1}, \nonumber\\
J_0 &=& V^{-1} D_0 V = V^{-1} \del_0 V - V^{-1} \, \alpha_0 \, V,
\label{8.7}
\eeqar
where $\alpha_0$ denotes a new gauge field.
It gauges $U$ on the right and $V$ on the left.
Further, notice that under $U \longleftrightarrow V^{-1}$,
\beqar
I_0 &\rightarrow& D_0 V^{-1} \, V = \del_0 V^{-1} \, V + V^{-1} \alpha_0 V =
- \left( V^{-1} \del_0 V - V^{-1} \alpha_0 V \right) = - J_0, \nonumber\\
J_0 &\rightarrow& U D_0 U^{-1} \,  =  U \del_0 U^{-1} \, V - U \alpha_0 U^{-1} =
- \left(  \del_0 U \, U^{-1} + U \alpha_0 U^{-1} \right) = - I_0, \nonumber\\
{\rm i.e.,}~~(I_0 , J_0) &\longleftrightarrow& (-J_0, -I_0).
\label{8.8}
\eeqar
The action is now taken as
\beqar
S &=& {i \over e^2} \int dt \int L, \nonumber\\
L&=& {1\over {\rm Im}\tau}  \Tr \left\{ \left[  I_0  \,\Bigl[  E (\tau - {\bar\tau})
+ i B (\tau + {\bar\tau}) \Bigr]\right]\right.  \left. +  \left[  J_0  \,\Bigl[  -B  (\tau - {\bar\tau})
+ i E (\tau + {\bar\tau}) \Bigr]\right]\right\}.
\label{8.9}
\eeqar
The overall $1/e^2$ is needed, but it is to be interpreted as some overall constant. While $\tau$ will undergo transformations under duality,
the overall $1/e^2$ is held constant. (It may be possible to absorb it into
the definition of the fields.)

The duality transformations correspond to modular transformations ($SL(2, \mathbb{Z})$)
of
$\tau$. These are generated by
\beq
{\rm a)~} \tau \rightarrow \tau + 1 ,\hskip .4in
{\rm b)~} \tau \rightarrow - {1\over \tau}.
\label{8.10}
\eeq
The first one is the same as $\theta \rightarrow \theta + 2 \pi$. Under this,
we  find from (\ref{8.5}),
\beq
W \rightarrow W - \int 2\, \Tr \left[\delta U U^{-1} {B \over 4 \pi}\right].
\label{8.11}
\eeq
The invariance of $e^{iS}$ is thus guaranteed by the quantization of the magnetic charge
$\oint (B/ 4\pi)$.
From $E (\tau + {\bar\tau})$ term in (\ref{8.9}), by a similar argument,
we will get quantization of
electric charge. Thus with the quantization of electric and magnetic
charges, the action  (or rather $e^{i S}$)
from (\ref{8.9}) is invariant under the
first of the modular transformations in (\ref{8.10}).

Now consider the second transformation in (\ref{8.10}).
Under $\tau \rightarrow - (1/\tau)$, we find
\beqar
{1\over {\rm Im}\tau} \Bigl[  E (\tau - {\bar\tau})
+ i B (\tau + {\bar\tau}) \Bigr] &\rightarrow& {1\over {\rm Im}\tau} \Bigl[  E (\tau - {\bar\tau})
- i B (\tau + {\bar\tau}) \Bigr], \nonumber\\
{1\over {\rm Im}\tau} \Bigl[  -B (\tau - {\bar\tau})
+ i E (\tau + {\bar\tau}) \Bigr] &\rightarrow& {1\over {\rm Im}\tau} \Bigl[  -B (\tau - {\bar\tau})
- i E (\tau + {\bar\tau}) \Bigr].
\label{8.12}
\eeqar
We can now accompany the second modular transformation by a transformation of the fields
defined as
\beq
E \longleftrightarrow B, \hskip .3in
U \longleftrightarrow V^{-1}.
\label{8.13}
\eeq
Using (\ref{8.12}) and this transformation,
\beqar
I_0 {1\over {\rm Im}\tau} \Bigl[  E (\tau - {\bar\tau})
+ i B (\tau + {\bar\tau}) \Bigr] &{\rightarrow}& I_0 {1\over {\rm Im}\tau} \Bigl[  E (\tau - {\bar\tau})
- i B (\tau + {\bar\tau}) \Bigr], \hskip .2in {\rm for~} \tau \rightarrow - 1/\tau\nonumber\\
&\rightarrow& (-J_0) {1\over {\rm Im}\tau} \Bigl[  B (\tau - {\bar\tau})
- i E (\tau + {\bar\tau}) \Bigr], \hskip .2in {\rm using~(\ref{8.13})}\nonumber\\
&=& J_0 {1\over {\rm Im}\tau} \Bigl[  -B (\tau - {\bar\tau})
+ i E (\tau + {\bar\tau}) \Bigr]\nonumber\\
J_0 {1\over {\rm Im}\tau} \Bigl[  -B (\tau - {\bar\tau})
+ i E (\tau + {\bar\tau}) \Bigr] &\rightarrow& 
I_0 {1\over {\rm Im}\tau} \Bigl[  E (\tau - {\bar\tau})
+ i B (\tau + {\bar\tau}) \Bigr].
\label{8.14}
\eeqar
Thus we have $L \rightarrow L$ in (\ref{8.9}).
The action is duality invariant if we do not transform the overall $1/e^2$ factor.

It is useful to rewrite the duality transformations.
The two generating ones are:
\begin{enumerate}
\item $\tau \rightarrow \tau +1$, $(E, B, U, V) \rightarrow (E, B, U, V) $
\item $ \tau \rightarrow - 1/\tau$, $(E, B, U, V) \rightarrow (B, E, V^{-1} , U^{-1}) $
\end{enumerate}
Compositions of these two will generate all the duality transformations.
Thus we have shown the duality invariance of $e^{i S}$ for the action
$S$ in (\ref{8.9}).

Our next step is to show that the electric and magnetic descriptions are obtained
as gauge choices for the new gauge field $\alpha_0$.
This can be done using the canonical analysis of the action.
To carry out this analysis, define
\beqar
\E &= &-{i\over 2 e^2 {\rm Im}\tau} \Bigl[  E (\tau - {\bar\tau})
+ i B (\tau + {\bar\tau}) \Bigr],\nonumber\\
\B&=& -{i\over 2 e^2 {\rm Im}\tau} \Bigl[  -B (\tau - {\bar\tau})
+ i E (\tau + {\bar\tau}) \Bigr].
\label{8.15}
\eeqar
The action is thus given by
\beqar
S &=& \int \Bigl[  -2  \Tr ( \del_0 U U^{-1} \E )  - 2 \Tr ( V^{-1} \del_0 V \B ) \nonumber\\
&&\hskip .3in
+ \alpha^a_0\, \left[ 2 \Tr ( U t_a U^{-1} t_b) \E_b - 2 \Tr(V^{-1} t_a V t_b) \B_b \right]
\Bigr].
\label{8.16}
\eeqar
Notice that the first two terms are similar to the co-adjoint orbit action.
But the fields
$\E$ and $\B$ are not restricted to the Cartan subalgebra, so 
the results are different from the case of the co-adjoint orbit actions.
In particular, the commutation rules are 
different. In fact, it is easy to see that the basic nontrivial
commutation rules are
given by
\beqar
[U_{ij} (x), U_{kl}(y) ] &=& 0, \nonumber\\
{}[\E_a(x), U(y)]&=& - t_a \, U(x)\, \delta (x, y), \label{8.17}\\
{}[\E_a(x), \E_b(y)]&=& i f_{abc} \E_c \, \delta (x, y). \nonumber
\eeqar
\beqar
[V_{ij} (x), V_{kl}(y) ] &=& 0, \nonumber\\
{}[\B_a(x), V(y)]&=& -  V(x)\, t_a\, \delta (x, y), \label{8.18}\\
{}[\B_a(x), \B_b(y)]&=&- i f_{abc} \B_c \, \delta (x, y). \nonumber
\eeqar
These relations also show that $2\, \Tr ( U t_a U^{-1} t_b) \,\E_b$ will carry out right translations on
$U$ and 
$2\, \Tr(V^{-1} t_a V t_b) \B_b$ will carry out left translations on $V$.

The equation of motion for $\alpha_0$ will give the Gauss law type constraint
\beq
2 \Tr ( U t_a U^{-1} t_b) \E_b - 2 \Tr(V^{-1} t_a V t_b) \B_b = 0.
\label{8.19}
\eeq
These are evidently first class constraints and so one can choose a gauge fixing
conjugate constraint.
Since the two terms in (\ref{8.19}) carry out right translations on $U$ and left translations on $V$, respectively,
there are two natural gauge choices: 
\begin{enumerate}
\item We can set $V = 1$ as the conjugate constraint. In this case,
the Gauss law (\ref{8.19}) shows that
$\B_a $ is equivalent to the generator of right translations on
$U$. The $V$-sector is eliminated and we have an ``electric description''.
\item We can set $U = 1$. In this case,
(\ref{8.19}) tells us that
$\E_a $ is equivalent to the generator of left translations on
$V$. The $U$-sector is eliminated and we have a ``magnetic description".
\end{enumerate}
The two dual descriptions appear as gauge choices now.
It is important that we can reduce to one or the other, so that the
degrees of freedom are as they are in the YM theory. As we said at the beginning, 
the YM theory should provide a complete accounting for all degrees of freedom.

There are also other gauge choices we can make, corresponding to mixed
electric-magnetic description.

Returning to the electric description, we can define the
operators $Q(\Theta )$ and $M(\Phi )$ by
\beqar
Q(\Theta ) &=& - \int d \Tr ( \Theta \E ) =
\oint \E_a \Theta^a, \nonumber\\
M(\Phi ) &=& \int \Tr ( \Phi\, U F^D U^{-1} ).
\label{8.20}
\eeqar
From the fact that $\E $ generates left translations on the group
element $U$, we see that the commutation relations
(\ref{mmcomm}), (\ref{chargecomm1}) and (\ref{chargecomm2}) are
easily obtained from (\ref{8.17}).
Further, notice that
\beq
\int \E = {1\over e^2} \int E  + {\theta \over 2 \pi} \int {B \over 4\pi}.
\label{8.21}
\eeq
The quantization of $\int \E$ implied by the fact it generates
group translations thus shows that the  electric charges have the additional
contribution, as expected from Witten's result.

While the canonical analysis and the commutation rules can be 
realized in the fully interacting quantum theory, there are difficulties with
the modular transformation.
Since $\tau$ involves the coupling constant, the full set of modular transformations cannot be realized
in general due to the running of the coupling constant.
This problem is avoided in theories 
with vanishing $\beta$-function, such as the ${\cal N} = 4$ supersymmetric theory or certain ${\cal N} = 2$ theories with judicious choice of matter content.

\section{Local Charges, Fluxes and Color Confinement}

So far, we have considered the edge effects defined on the sphere at
spatial infinity, namely, with $r \rightarrow \infty$. In this section, we 
comment on the case of the local algebra of observables defined
on a bounded set. 

On the spatial slice, let ${\cal O}$ be a bounded open set and ${\cal A}({\cal O})$ the local algebra of observables. They are gauge invariant operators of local fields smeared with Lie algebra valued test functions $f$ with support $\subset {\cal O}$, i.e., operators such as
\begin{equation}
F^2 (f) = \int_{\cal O} d^{N-1}x \, \text{Tr}\Big(\left[F^2(x) \right]f(x) \Big).
\label{6.1}
\end{equation}
Suitable regularization may be needed to make these well-defined; we can also consider their exponentiated forms to get bounded operators. 

Now consider $Q(\Theta)$ and $M(\Phi)$ where the test functions are supported in $\overline{\cal O}$, the closure of ${\cal O}$. It is then clear that they commute with the local observables ${\cal A}({\cal O})$ of operators localized in ${\cal O}$.
For test functions $\Theta$ with support in open sets $V \subset {\cal O}$, $Q(\Theta)$ is an integral of the Gauss law and hence vanishes. That is also the case for $M(\Phi)$ iff $\text{Supp } \Phi = V \subset {\cal O}$, as shown by an integration by parts and use of the Bianchi identity.
Thus the observables are in ${\cal A}({\cal O})$, with the two-sided ideals generated by the Gauss law and $M(\Phi)$ with test functions having support in $V$ quotiented out. 

But the situation for  $Q(\Theta)$ and $M(\Phi)$ with support $\overline{\cal O}$ for $\Theta$ and $\Phi$ is different. 
They commute with the Gauss law and are gauge-invariant. They also commute with ${\cal A}({\cal O}')$ and can be regarded as additional operators in ${\cal A}(\overline{\cal O}) \supset {\cal A}({\cal O})$. 

The centre of the algebra generated by $Q(\Theta)$ and $M(\Phi)$ obviously also commutes with ${\cal A}({\cal O})$.
The algebra generated by $Q(\Theta)$ and $M(\Phi)$ define the edge observables of ${\cal A}(\overline{\cal O})$. It is non-abelian and has the defining relations 
\begin{eqnarray}
[Q(\Theta_1), Q(\Theta_2)] &=& Q([\Theta_1, \Theta_2]), \nonumber\\
\protect{[}Q(\Theta), M(\Phi)] &=& M([\Theta, \Phi]), \label{6.2}\\
\protect{[}M(\Phi_1), M(\Phi_2)] &=& 0. \nonumber
\end{eqnarray}
These are identical to (\ref{mmcomm}), (\ref{chargecomm1}) and (\ref{chargecomm2}), but now for the case of a bounded set $\overline{\cal O}$. We will also refer to the group action generated by these operators
as the generalized sky group.

It is also useful to consider the action of these operators on states.
Consider first the vector states on ${\cal A}({\cal O})$ obtained from the vacuum density matrix $|0\rangle \langle 0|$ and the GNS construction. These vectors have zero charge and no edge states, so that $Q(\Theta)$ and $M(\Phi)$ acting on them give zero. The vacuum sector has no charges or fluxes. 

Consider next a complex quantum scalar field $\phi_\alpha$ on which the generalized sky group of ${\cal O}$ generated by (\ref{qontheta})
has an adjoint action. We next smear it with a test function $f^\alpha$ localized in ${\cal O}$ to get
\begin{equation}
\phi(f) = \int_{\overline{\cal O}} d^{N-1}x \,\phi_\alpha (x) f^\alpha(x).
\label{6.3}
\end{equation}
We then consider the vector 
\begin{equation}
\phi(f) |0\rangle := |\phi(f) \rangle.
\label{fstate}
\end{equation}
For an interacting field $\phi$, the normalization of (\ref{fstate}) to 1, leading to a density matrix
\begin{equation}
\omega_f = |\phi(f) \rangle \langle \phi(f) |
\label{6.5}
\end{equation}
can be rather problematic.
For this reason, we take $\phi$ be a free, in- or out-field with $|0\rangle$ 
as the corresponding vacuum. This suffices to illustrate the point
we are making.
For free fields, we can normalize $\omega_f$, that is, ensure that 
$\omega_f(\mathbb{1}) =1$
by setting 
\begin{equation}
\int_{\overline{\cal O}} d^{N-1}x \, \bar{f}(x) f(x) =1.
\label{fnorm}
\end{equation}
With (\ref{fnorm}), $\omega_f$ is a state on ${\cal A}({\cal O})$. Therefore, by the GNS construction, we can get a representation of 
${\cal A}({\cal O})$ on a Hilbert space ${\cal H}({\cal O})$. 
(We can do this even though $| \phi(f)\rangle$ is not annihilated by `small' gauge transformations.)

The Gauss law operators with test functions vanishing on ${\overline{\cal O}} / {\cal O}$ vanish on states as implied by the preceding discussion. But the generalized sky group of $\overline{\cal O}$ can act non-trivially on states including vector states $\omega_f({\cal O})$. So we can extend this representation to the extension of ${\cal A}({\cal O})$ which includes the group algebra of the generalized sky group. This is the representation built from a general vector state.

Since the generalized sky group acts nontrivially on these vector
states, clearly they are the starting point towards a definition
of color confinement.
In a conventional treatment of first class constraints, such as those which arise in gauge theory Lagrangians, the constraints {\it for all test functions} are set to zero on the allowed state vectors. The observables are also required to be gauge-invariant. No constraints are placed on the test functions for the Gauss law. In this approach, since both observables and states are color singlets, there is no notion of color left: there is no observable operator with color whose zero expectation value will signal color confinement.

The imposition of Gauss law on states with no constraints on the test functions eliminates all charged states.
The more general structure is to impose the Gauss law on physical states
with test functions which vanish at spatial infinity.
The operators  $Q(\Theta )$ with a constant nonzero value for $\Theta (x)$
as $r \rightarrow \infty$ define the charge algebra with states of nonzero charges transforming appropriately under the action of
$e^{i Q (\Theta )}$.
Allowing for $\Theta$ to be asymptotically a function on $S^{N-2}$
as in
\cite{Balachandran:2013wsa,Balachandran:2018cgw,Balachandran:2018jwf} 
enhances this framework. Since
only the smeared Gauss law with test functions vanishing at infinity is required to vanish on vector states,  the latter can in general
transform nontrivially under the sky group, which includes the global $G$.
Thus colored states are not {\it a priori} ruled out at the level of
the gauge algebra, and 
color confinement has to be understood dynamically as the statement
that the expectation values of the Hamiltonian in the colored states are infinite. In other words, colored states are not in the domain of the Hamiltonian.
While the expectation value of the Wilson loop or Polyakov loop
operators can serve as a diagnostic of confinement in certain 
circumstances (e.g. in gauge theories invariant under the action of the
center of the group $G$), the
proper definition of confinement has to be phrased in terms of the domain
of the Hamiltonian.

This scenario is applicable in the even more general approach suggested in this paper: we have required only the observables ${\cal A}$ to be gauge-invariant, then from any state on ${\cal A}$, gauge-invariant or not, using the GNS construction, we can recover a $*$-representation of ${\cal A}$. We can then extend this representation by including the group for charges and fluxes, $Q(\Theta)$ and $M(\Phi)$, as well. 

We cannot claim that such a treatment of states is new. The DHR analysis \cite{Haag:1992hx} where the observables ${\cal A}$ are invariant under a group $G$, which from the context, we can infer is the global symmetry group, is treated in this manner. It has also appeared in the constraint analysis of the action for gravity, under the guise of ``frozen" formalism. The algebra ${\cal A}$ is then the algebra of diffeo-invariant observables with no notion of time. A $*$-representation of ${\cal A}$ then defines the quantum theory. This approach was applied in \cite{Balachandran:2017jha} to quantize the relativistic action (the coadjoint action) of a point particle with reparametrization invariance. This treatment does not require gauge fixing and gives for ${\cal A}$ the Poincar\'e algebra with its centre generated by the mass and spin Casimirs fixed. So ${\cal A}$ has only one representation, namely that with fixed mass and spin, just as one wants.

%%%%%%%%%%%%%%%%%%%%%%%%%%%%%%%%%%%%%
%%%%%%%%%%%%%%%%%%%%%%%%%%%%%%%%%%%%%
%%%%%%%%%%%%%%%%%%%%%%%%%%%%%%%%%%%%%
\section{Discussion}

In this paper, we have considered the operator $Q(\Theta )$ 
and its magnetic dual $M(\Phi )$, which together define the superselection sectors of a gauge theory. 
The algebra of these operators, which we refer to as the Lie algebra
of the generalized sky group, is given in
(\ref{mmcomm}), (\ref{chargecomm1}), (\ref{chargecomm2}).
In the nonabelian case, the superselection also implies that
$\G (G)$, the group of maps from the spatial boundary to the
group $G$, is spontaneously broken.
For the four-dimensional gauge theory, we have also obtained
a duality-invariant action, which allows one to go from an
electric description of the generalized sky group to a magnetic
description, realizing the $SL(2, \mathbb{Z})$ modular
transformations on the complex coupling constant $\tau$.
We have also commented on the definition of charges and confinement
in terms of the representations of the operators $Q(\Theta )$ and
$M(\Phi )$.

We will close this section with a few comments on
some additional mathematical
structures associated with $Q(\Theta)$ and $M(\Phi)$,
even though their physical implications are 
still unclear.

 \noindent\underline{Drinfel'd Double}

The Drinfel'd double arises from a Hopf algebra $H$ and the linear functionals $H^*$ on $H$ 
with values in $\mathbb{C}$ with its induced Hopf algebra structure. Then $H \ltimes H^*$, the crossed 
product of $H$ with $H^*$ is also a Hopf algebra, the Drinfel'd double of $H$.

Another manner in which it can arise is as follows. Let $G$ be a group with an action on another group $N$. Now $\mathbb{C}G$ and $\mathbb{C}N$ are canonically Hopf algebras and the action of $G$ on $N$ induces a Hopf algebra action of $\mathbb{C}G$ on $\mathbb{C}N$. The crossed product $\mathbb{C}G \ltimes \mathbb{C}N$ is also a Hopf algebra, the Drinfel'd double associated with $G$ and $N$. 

In our case, we can also give a pairing of $Q(\Theta)$ and $M(\Phi)$ so that they are mutually dual. This is given by
\begin{equation}
\langle Q(\Theta), M(\Phi) \rangle = \int_{S^{N-2}} d \Omega_{S^{N-2}}(\hat{n}) \text{Tr } (\bar{\Theta}^\infty \Phi^\infty) (\hat{n}).
\label{qmpairing}
\end{equation}
The implications of this structure are not clear. 

But there is also another approach pioneered by Doplicher, Haag and Roberts \cite{Haag:1992hx}. It is based on the Cuntz algebra $O_d$ and the endomorphisms on the representations of these electric and magnetic groups. We can get the Cuntz algebra for gauge theories as follows.
Again, we introduce a basis $t_a$ for $\underline{G}$ with normalization $\text{Tr } t_a t_b = \half \delta_{ab}$ and correspondingly, test functions $\Theta = \Theta^a t_a, \Phi = \Phi^a t_a$. That gives 
\begin{equation}
Q(\Theta) := \chi^a(\Theta) t_a, \quad M(\Phi) := \rho^a(\Phi) t_a,
\label{7.1}
\end{equation}
where $\chi^a$ and $\rho^a$ are complex-valued test functions of their arguments. Under the adjoint action of $G$, $\chi^a(\Theta)$ and $\rho^a(\Phi)$ can be chosen to transform by unitary matrices. Thus for example
\begin{equation}
G \ni g \rhd \chi^a(\Phi) \rightarrow \chi^b(\Phi) D_{ba}(g), \quad \text{where }D^\dagger D = DD^\dagger = \mathbb{1}.
\label{7.2}
\end{equation}
We can then consider the combination
$\chi(\Theta^a) + i \rho(\Theta^a)$
and its polar decomposition 
\begin{equation}
\chi^a(\Theta) + i \rho^a(\Theta) = \psi_a \,\,|| \chi^a(\Theta) + i \rho^a(\Theta) ||.
\label{7.3}
\end{equation}
Following Fredenhagen's review \cite{fredenhagen}, it is possible to show that $|| \chi^a(\Theta) + i \rho^a(\Theta) ||$ is independent of $a$ and that
\begin{eqnarray}
\psi_a \psi_b^* &=& \delta_{ab} \mathbb{1}, \label{cuntz1}\\
\sum \psi_a^* \psi_a &=& \mathbb{1} \label{cuntz2} .
\end{eqnarray}
These relations define the Cuntz algebra of partial isometries $O_d^0$  and after its completion in the $C^*$-norm, the Cuntz algebra $O_d$. Further, as may be expected,
\begin{equation}
e^{i Q(\Theta)} \psi_a e^{-i Q(\Theta)}  = \psi_b \, \left[U(e^{i Q(\Theta)})\right]_{ba}.
\label{7.4}
\end{equation}

With these preliminary statements, we can
define the Drinfel'd double. Functions on ${\cal G}(G)$ have the basis 
\begin{equation}
f_{a_1,\cdots a_k;a'_1 \cdots a'_k} = \psi_{a_1} \psi_{a_2} \cdots \psi_{a_k} |0 \rangle \langle 0| \psi^*_{a'_k} \psi^*_{a'_{k-1}} \cdots \psi^*_{a'_1}.
\label{7.5}
\end{equation}
At a point $e^{i Q(\Theta)} $ of the group ${\cal G}(G)$, it has the value 
\begin{equation}
f_{a_1,\cdots a_k;a'_1 \cdots a'_k} (e^{i Q(\Theta)}) = \langle 0| \psi^*_{a'_k} \psi^*_{a'_{k-1}} \cdots \psi^*_{a'_1}e^{i Q(\Theta)} \psi_{a_1} \psi_{a_2} \cdots \psi_{a_k} |0 \rangle.
\label{7.6}
\end{equation}
The group ${\cal G}(G)$ acts on these functions by left translation:
\begin{eqnarray}
e^{i Q(\Theta)} \rhd f_{i_1,\cdots i_k;i'_1 \cdots i'_k} &\rightarrow& e^{i Q(\Theta)} f_{i_1,\cdots i_k;i'_1 \cdots i'_k}, \\
(e^{i Q(\Theta_1)}f_{i_1,\cdots i_k;i'_1 \cdots i'_k}) (e^{i Q(\Theta_2)}) &=& f_{i_1,\cdots i_k;i'_1 \cdots i'_k} (e^{-i Q(\Theta_1)} e^{i Q(\Theta_2)}).
\end{eqnarray}
So we have the structure of a Drinfel'd double.

\noindent\underline{The Manin Triple}

The Manin triple ${\cal L}$ is a Lie algebra with two subalgebras ${\cal L}_1$ and ${\cal L}_2$ with ${\cal L}_1 \cap {\cal L}_2 = \{0\}$ and ${\cal L} = {\cal L}_1 \oplus {\cal L}_2$, and a metric $g$ for which ${\cal L}_1$ and ${\cal L}_2$ are isotropic subalgebras:
\begin{equation}
g(l_i',l_i) = 0, \quad l_i,l_i' \in {\cal L}_i, \quad  i=1,2.
\end{equation}
We can see that the pairing (\ref{qmpairing}) implies a Manin triple for the gauge theory. Thus consider a Lie algebra ${\cal L}$ defined as
${\cal L} = \text{Span}( Q(\Theta), M(\Phi))$ and the two subalgebras ${\cal L}_1 = \text{Span}( Q(\Theta))$, ${\cal L}_2 = \text{Span}(M(\Phi))$. We have given their commutators in (\ref{mmcomm}), (\ref{chargecomm1}) and (\ref{chargecomm2}). With the pairing in (\ref{qmpairing}),
we can now define the metric $g$ on ${\cal L}$:
\begin{eqnarray}
g(Q(\Theta),Q(\Theta)) &=& g(M(\Phi),M(\Phi)) =0, \\
g(Q(\Theta),M(\Phi)) &=& g(M(\Phi),Q(\Theta)) = \int_{S^{n-2}} d \Omega_{S^{n-2}} (\hat{n}) \text{Tr } (\overline{\Theta}^\infty \Phi^\infty).
\end{eqnarray}
It is clear from these equations that ${\cal L}_1$ and ${\cal L}_2$ are isotropic subalgebras.

In finite dimensions, say $d$, it is such a metric which has $O(d,d)$ symmetry with the group $O(d,d,\mathbb{Z})$ underlying $T$-duality. What happens here, in infinite dimensions, remains to be understood.

 \noindent\underline{A comment on magnetic monopoles}

Classically, magnetic monopoles can be identified by examining 
$\Phi^\infty$, the test function for $M(\Phi )$ at infinity. Thus consider $G=SU(2)$ and let $\Phi^\infty(\hat{n})$ lie on a non-zero orbit of this group. Since $\Phi^\infty(\hat{n})$ is now $SU(2)$ Lie algebra valued, and the stability group of a point here is $U(1)$, its possible values are in $S^2$. But $S^{N-2}$ is also $S^2$ for $N=4$. So $\Phi^\infty(\hat{n})$ is a map from $S^2$ to $S^2$. Such maps with winding number $n$ are associated with the infinite number of test functions for monopole charge $n$. Then the classical observable $M(\Phi^\infty)$ is a measure of the magnetic charge $n$. 
In the quantum theory, $M(\Phi^\infty)$ is an operator. So we need a {\it quantum state} with expectation value approximating the classical answer. 
This will require a quantum operator which acting on, say, the vacuum of the theory can create such a state. As we saw earlier, an 't Hooft-Polyakov monopole state twisted by the the Chern-Simons term is a semi-classical candidate. But this issue requires further investigation.

\noindent\underline{On the local Cuntz algebras}

The Cuntz algebras (\ref{cuntz1}), (\ref{cuntz2}) can be adapted to local regions by restricting the supports of 
$\Theta^j$ to ${\cal O}$. It is such local algebras that arose in the papers of DHR \cite{Doplicher:1969tk}-\cite{Doplicher:1990pn}.

The vector states transforming by the sky group are then readily constructed
as
\begin{equation}
\psi^*_{a_1} \psi^*_{a_2} \cdots \psi^*_{a_k} |0 \rangle.
\end{equation}
They have unit norm by (\ref{cuntz1}). They can replace (\ref{fstate}) to get a representation of the sky group and ${\cal A}({\cal O})$.

\noindent {\bf Acknowledgments:} AFRL acknowledges financial support from the Faculty of Sciences of Universidad de los Andes through project INV-2019-84-1833. The work of VPN was supported in part by the U.S. National Science Foundation Grants No. PHY-2112729 and No. PHY-1820271 and by PSC-CUNY awards.

%%%%%%%%%%%%%%%%%%%%%%%%%%%%%%%%%%%%
%%%%%%%%%%%%%%%%%%%%%%%%%%%%%%%%%%%%
%%%%%%%%%%%%%%%%%%%%%%%%%%%%%%%%%%%%

%%%%%%%%%%%%%%%%%%%%%%%%%%%%%%%%%%%%
%%%%%%%%%%%%%%%%%%%%%%%%%%%%%%%%%%%%
%%%%%%%%%%%%%%%%%%%%%%%%%%%%%%%%%%%%
\end{document}